\documentclass[aip,chaos,graphicx,twocolumn,10pt]{revtex4-1}
\usepackage{graphicx}
\usepackage{color}

\begin{document}


\title{Competing structures in a minimal double-well-potential model of condensed matter} 



\author{Julyan H. E. Cartwright}
\email{julyan.cartwright@csic.es}
\affiliation{Instituto Andaluz de Ciencias de la Tierra, IACT-CSIC, 18100 Armilla, Granada, Spain}
\affiliation{Instituto Carlos I de F\'{\i}sica Te\'orica y Computacional, Universidad  de Granada, 18071 Granada, Spain}
\author{Bruno Escribano}
\email{bescribano@cab.inta-csic.es}
\affiliation{Centro de Astrobiolog\'ia (CAB), CSIC-INTA, 28850 Torrej\'on de Ardoz, Madrid, Spain}
\author{S\'andalo Rold\'an-Vargas}
\email{sandalo@ugr.es}
\affiliation{Department of Applied Physics, Faculty of Sciences, Universidad de Granada, 18071 Granada, Spain}
\author{C. Ignacio Sainz-D\'{\i}az}
\email{ci.sainz@csic.es}
\affiliation{Instituto Andaluz de Ciencias de la Tierra, IACT-CSIC, 18100 Armilla, Granada, Spain}


\date{\today}

\begin{abstract}
The microscopic structure of several amorphous substances often reveals complex patterns such as medium- or long-range order, spatial heterogeneity, and even local polycrystallinity. To capture all these features, models usually incorporate a refined description of the particle interaction that includes an \textit{ad hoc} design of the inside of the system constituents, and use temperature as a control parameter. We show that all these features can emerge from a minimal athermal two-dimensional model where particles interact isotropically by a double-well potential, which includes an excluded volume and a maximum coordination number. The rich variety of structural patterns shown by this simple geometrical model apply to a wide range of real systems including water, silicon, and different amorphous materials.
\end{abstract}


\maketitle 

\textbf{Crystalline solids are characterized by their periodic internal structure. They can be experimentally studied using diffraction patterns and numerically modeled using their unit cell lattice. 
On the other hand, amorphous materials have no apparent periodicity, they are often described as having disordered or randomized internal structures, but they still display predictable properties such as density or conductivity, and they can exhibit phase transitions between several amorphous states. 
Dynamic models for such materials are often built using central potentials that are purposely designed to reproduce experimental observations. 
The increasing complexity of such potentials can often disregard the fundamental cause of polyamorphism. 
In this work we propose a minimalist geometrical model using the simplest double-well potential that can still display pattern formation in an amorphous solid. 
Our results show that a double-well potential may be sufficient for the emergence of long-range order, polycrystallinity and polyamorphism.
}

\section{Introduction}
Much research, both laboratory experiments and numerical simulations, is taking place on substances that show multiple liquid and amorphous solid phases. These include water \cite{kim2020,rosu2023}, silicon \cite{morishita2004high,ganesh2009liquid,sastry_silicon,treacy2012}, carbon \cite{togaya1997,glosli1999liquid,sachan2019}, phosphorus \cite{katayama2000,morishita2001liquid,monaco2003nature},  
silica \cite{lacks2000first,trave2002pressure}, triphenyl phosphite \cite{kurita2004}, carbon dioxide \cite{bai2018}, and calcium carbonate \cite{cartwright2012_2}. Noted phenomena are long- or medium-range order despite the system being amorphous, spatial heterogeneity, both dynamic and static, and local orientation \cite{lan2021}. To rationalize these features, which are common to many amorphous solid materials, it has been proposed that a particle pair potential having the form of a double well may be the generic mechanism for this rich phenomenology, termed in general polyamorphism \cite{buldyrev2002,wilding2006}. In particular, confined and interfacial two-dimensional materials are being investigated in theory, simulations, 
and experiments not only for their wide range of applications but also because of their fundamental role in clarifying different surface phenomena \cite{li2021}. In this work we focus on two dimensions and demonstrate that long-range order that extends further than the typical liquid order, static spatial heterogeneity, and local orientation emerge from a minimal model with a double well and isotropic interaction.

More complex approaches with a double-well potential, e.g., models based on molecular thermodynamics, have  shown water-like anomalies
and/or a polyamorphic transition \cite{franzese2001generic,malescio2007complex,franzese2007differences,Rosu-Finsen2023,Mollica2022}. 
Core-softened potentials such as the Hemmer-Stell-Jagla model
display two phase transitions and anomalous-waterlike properties
\cite{hemmer1970fluids,stell1972phase,jagla1999core}. Likewise, previous work with a minimal model of water, the so-called Mercedes-Benz model that incorporates directional bonding by representing water molecules as two-dimensional Lennard-Jones disks with three equivalent hydrogen-bonding arms disposed at 120 degrees, displays a first-order phase transition between a crystalline phase and a high-density amorphous phase as well as a reversible transformation between two amorphous structures of high and low density \cite{cartwright2012}.

Here we investigate an even simpler model in two dimensions containing only a double well with no directionality in the particle bonding mechanism and without thermodynamical considerations. 

The choice of a double-well potential differs from previous mechanisms leading to amorphous structures. In our work a given particle can create bonds at different distances.  {
This approach represents a simple way to avoid crystallization, thereby producing amorphous structures. Typically, the common mechanism to avoid crystallization in canonical models of glass-forming liquids has been based on the use of particles of different size, in particular binary mixtures constituted by A and B particles. Such binary models necessitate several parameters, not only the three different interaction ranges associated with AA, AB, and BB interactions but also three different depths associated with their respective interaction wells (see for instance Ref. \cite{kob-andersen_2D} for Lennard-Jones interactions). Our double-well approach reduces  the number of parameters while still producing amorphous structures.
}

\begin{figure}[tb]
\includegraphics[width=1.0\linewidth]{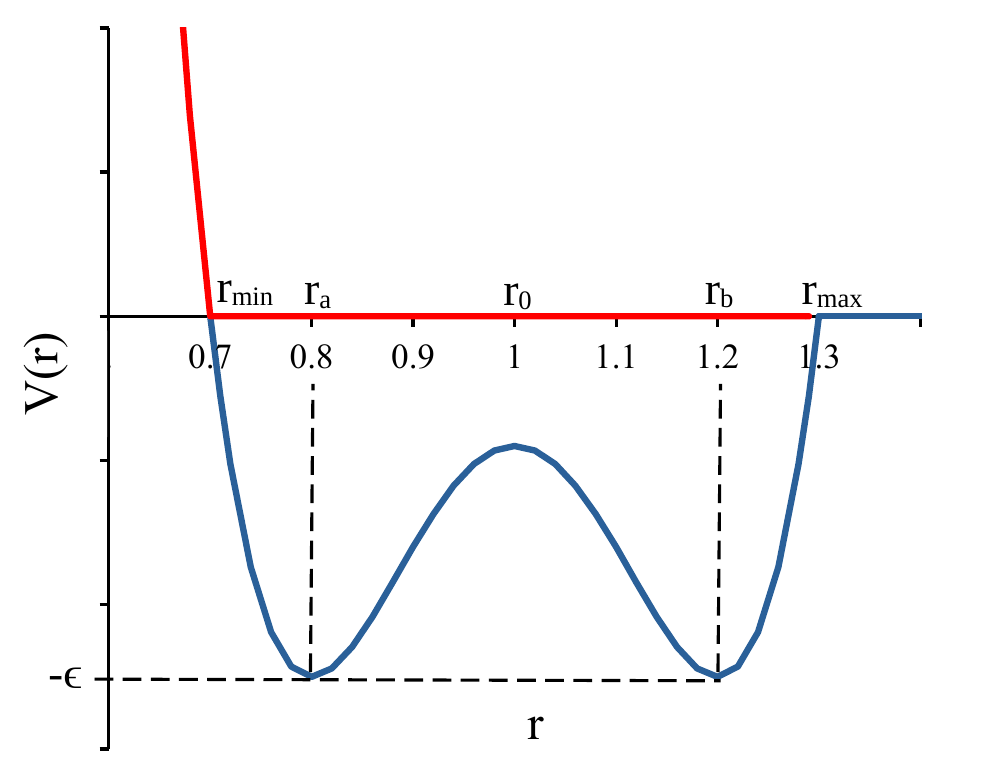}
\caption{\textbf{Double-well potential with excluded volume}. The blue line represents the potential $V(r)$ for bonded interactions with a maximum coordination limit, $n$. The red line represents the unbonded interaction potential $V_{unbonded}(r)$ in the form of an excluded volume.}
\label{fig_1_potential}
\end{figure}

\section{Methods}
With that objective in mind, we simulate the growth of a two-dimensional amorphous solid material by depositing particles on a plane surface. Particles are deposited one at a time randomly with uniform probability on the simulated plane, and can interact with each other forming permanent bonds by proximity which are not directionally constrained. In between depositions, we allow the system to relax and stabilize using a second-order Euler integrator for the equations of motion, for which we assume under-damping to avoid numerical instabilities.
Typically, we allowed 10 time-steps of the Euler integrator between depositions of new particles. We did not study the effect of longer relaxations which would simulate slower depositions.

We use a double-well potential for the interaction between a pair of bound particles separated a distance $r$,

\begin{equation}
V(r)= \bigg\{ \begin{array}{lr} \frac{1}{2}  k  (r-r_a)^2  (r-r_b)^2 - \epsilon & \text{if } r \leq r_{max}\\ 0 & \text{if } r > r_{max},
\end{array} 
\end{equation}

where $r_a$ and $r_b$ are the equilibrium distances for the two wells; Fig.~\ref{fig_1_potential}. For simplicity, we choose $k=1$ and model the particles as point masses with mass $m=1$, defining the unit of length $r_0={(r_a+r_b)/2=1}$ as the distance of the potential barrier, {which symmetrically separates the two wells}. In addition, we introduce $r_{max}$ as the maximum distance for bond formation, which is necessary for avoiding the divergent asymptotic behavior of the potential, and set $\epsilon=\frac{1}{2}  k  (r_{max}-r_a)^2  (r_{max}-r_b)^2$. Hence we have particle bond lengths not very far from $r_a$ and $r_b$. In this respect, we also impose bonds to be irreversible, in the sense that, once two particles are bonded, they could interact by $V(r)$ even for distances greater than $r_{max}$, although in the explored systems this situation is marginal. 

We control the number of bonds per particle by imposing a maximum coordination number, $n$; each particle can bond with at most $n$ neighbors. 
This is imposed during the neighbor search algorithm, which is executed at every time-step of the Euler method for the integration of the equations of motion. By altering the maximum coordination number, we obtain different competing structures without appealing to thermodynamic considerations based on the interplay between entropically and energetically favorable structures \cite{gel_inverso}. {The maximum coordination number we introduce is, in our view, the simplest manner to incorporate valence; we neither impose bond directionality nor introduce anisotropy in the internal geometrical description of our particles. However, as we will show, our model, with a given value for the maximum coordination number and only radial interactions, still reveals clear local orientations between particles as well as polycrystalline structures.
} 

Once a particle reaches $n$ bonds it becomes saturated and does not create additional bonds with other particles in its vicinity. In this respect, we also allow non-bonded repulsive interactions for $r \leq r_{min}$, given by the potential $V_{unbonded}(r)= V(r)$ if  $r \leq r_{min}$ and $V_{unbonded}$ = 0 if  $r > r_{min}$ (red line in Fig.~\ref{fig_1_potential}). This repulsive interaction represents a form of excluded volume, keeping non-bonded particles from a significant overlap between each other for $r \le r_{min}$ while moving during the system relaxation after each deposition. {Without introducing an excluded volume, our deposition mechanism would lead to spurious structures in which non-bonded particles could be arbitrarily close.} In this respect, we also note that when a new particle is randomly deposited at a distance $r \leq r_{min}$ from a previously deposited particle, we remove that new particle and try another random deposition.

\begin{figure}[tb]
\centering
\includegraphics[width=1.0\linewidth]{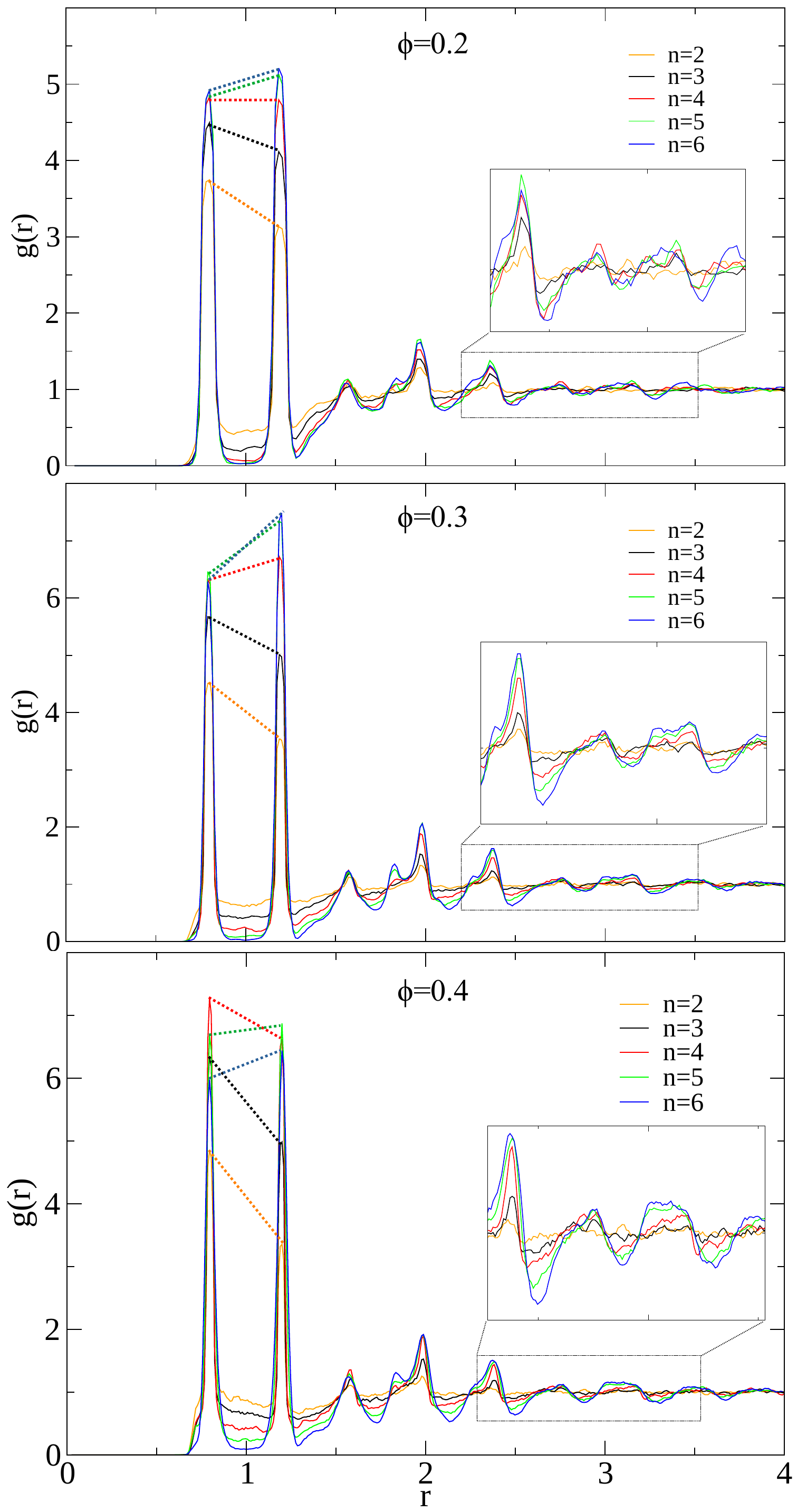}
\caption{\textbf{Radial distribution function}  $g(r)$ as a function of $\phi$ and $n$. Dashed lines mark the height of the two main peaks,  at $r_a$ and $r_b$, for clarity. Inset show a zoom with a detail of the medium-range order for $r>2$.}
\label{gr}
\end{figure}

We set the distances in our model as $r_{min}=0.7$, $r_a=0.8$, $r_b=1.2$ and $r_{max}=1.3$.  With these choices we: i) place the two wells, as well as $r_{min}$ and $r_{max}$, symmetrically with respect to the potential barrier (thus effectively reducing the complexity of our model), ii) impose a ratio $
{r_{min}/r_a}=0.875$ similar to the one associated to a Lennard-Jones interaction between the Lennard-Jones particle diameter and the corresponding  Lennard-Jones minimum \cite{lennard-jones1931}, and iii) allow any two bonded particles to cross the barrier in either sense, and jump from one well to the other during the relaxation after depositions. To investigate the effect of the occupied surface, we introduce a packing-fraction parameter,
$
\phi={N \pi (r_{min}/2)^2/L^2},
$
where $L=100$ is the fixed side length of our square simulation box, with periodic boundary conditions, and $N$ the total number of particles, which we alter to explore different packing fractions. {Since our study focuses on  properties in the bulk phase, we use periodic boundary conditions as an efficient method to avoid undesirable artifacts due to border interactions. To this effect, we impose a sufficiently large ratio between $L$ and $r_{max}$\cite{Allen-Tildesley,Fenkel-Smit}. Moreover, the values of $L$ and $N$ we use in this work were previously tested with other smaller values in order to remove finite-size effects and
assure the consistency of our results.}. We present results for $\phi=0.2$, $0.3$, and $0.4$ --- $N=5197$, $7795$, and  $10394$ --- that is, from a dilute system to a packing fraction close to the random packing fraction of equal non-overlapping disks \cite{hinrichsen1990}. To isolate the structural effect of the maximum coordination number, we cover, for any given $\phi$, a wide range of $n\in \{2,3,4,5,6\}$. We analyze 40 independent simulations for each pair $(\phi, n)$. {We tested that this number of simulations is sufficient to give consistent statistical results for all the coordination numbers and packing fractions explored}. 

\begin{figure}[tb]
\centering
\includegraphics[width=1.0\linewidth]{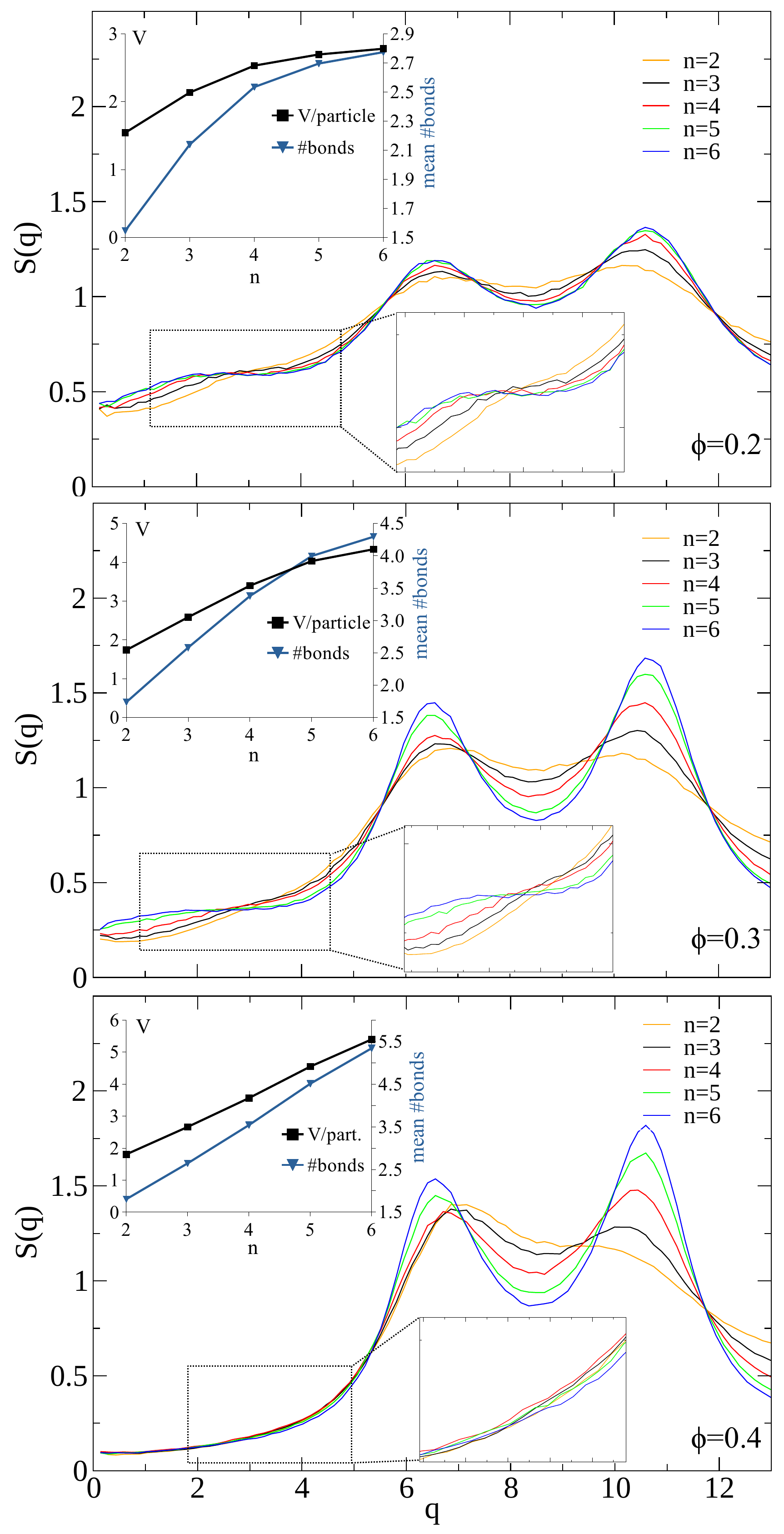}
\caption{\textbf{Structure factor}  $S(q)$ as a function of $\phi$ and $n$. Insets show, for each $\phi$, the energy per particle, $V/$particle, as well as the mean number of bonds per particle, as a function of $n$. Insets show a zoom of the long-range order manifest at low $q$.}
\label{sq}
\end{figure}

\section{Results and discussion}
We begin analyzing the system structure by looking at the radial distribution function $g(r)$~\cite{Hansen-McDonald}  as a function of $\phi$ and $n$, Fig.~\ref{gr}. Notice the emergence of two main peaks placed at $r_a$ and $r_b$, the two potential wells, whose height increases with $\phi$, with independence of $n$. Second, for any given $\phi$, $g(r)$ shows a more obvious (amorphous) order when increasing $n$. This is again manifested through the height of the two main peaks. More interestingly, we observe how for low $n$ ($=2,3$), and for any given $\phi$, the first of the two main peaks, placed at $r_a$, is more likely than the second peak placed at $r_b$; see the yellow and black dashed lines in Fig.~\ref{gr}. However for high coordination numbers, $n=5$ and $6$, the second peak is more likely than the first; see the green and blue dashed lines in Fig.~\ref{gr}. In simple terms, for high coordination numbers particles tend to bond to their neighbors at larger distances. Presumably, this more frequent bonding with the second well has a clear effect on the medium-range order for $n=5$ and $6$, especially at $\phi=0.3$ and $0.4$, where the medium-range system structure is manifested through well developed peaks for $r>2$; see insets in Fig.~\ref{gr}. The phenomenology shown by $g(r)$ already displays the versatility of our model, which produces a medium-range order that extends further than the typical liquid structure as well as a means to control the first-neighbor structure by controlling $n$. This non-trivial structure can be connected with previous---and more complex---models and experiments. For instance: i) Koga et al.~\cite{koga2000} analyzed a quasi-2D amorphous water model by molecular-dynamics simulations using a TIP4P force field which incorporates a refined description of the internal geometry of the water molecules. They also found two main peaks and a medium-range order similar to those shown by our model, ii) similar profiles for $g(r)$ have been reported recently by Negi et al.~\cite{negi2022} for a two-dimensional ice model using a sophisticated density-functional theory approach, iii) Treacy et al.~\cite{treacy2012} studied the amorphous phase of silicon by electron-diffraction experiments, finding a $g(r)$ profile consistent with our results: two main peaks and additional small peaks at longer distances; in particular, they found a coordination number of 3.8 for a narrow 3D system confined between two walls.

This non-trivial order is manifested more clearly when looking at the structure factor $S(q)$~\cite{Hansen-McDonald}, Fig.~\ref{sq}. First, we observe two main peaks around $q$ $\approx$ 6 and 11, in accordance with the two main peaks shown by $g(r)$ for the two potential wells. As in $g(r)$, these two peaks are more obvious upon increasing $\phi$ and $n$.  Long-range order is now seen at low $q$, at distances even longer than those manisfested by $g(r)$. Again,  this long-range order is obvious for high coordination numbers and, especially, at intermediate $\phi$. Thus, for $q <$ 4 and $\phi=0.3$, the system develops a plateau at high $n$ ($= 5, 6$) which tends to disappear at low coordination numbers, $n = 2$, $3$; see insets in Fig.~\ref{sq}. This plateau, particularly apparent around $q$ $\approx$ 2, corresponds to distances of the order of $r \geq 3$, and can be interpreted as the emergence of long-range spatial heterogeneities in the particle density. The long-range order indeed extends quite far ($r \approx 6$, $q \le 1$), where the limit of $S(q)$ when $q\rightarrow{0}$ produces higher values upon increasing $n$, making the system more compressible, in analogy with systems controlled by temperature~\cite{Hansen-McDonald}. 
The plateau disappears at $\phi=0.4$ for any $n$ and the system recovers a liquid-like spatial homogeneity, despite maintaining an obvious local medium-range order. At these high packing fractions ($\phi=0.4$) the long-range spatial heterogeneities shown by $S(q)$ for $\phi=0.3$ disappears presumably due to the absence of empty space. $S(q)$ is directly measurable in scattering experiments~\cite{Hansen-McDonald} and, in particular, the structure factor plots we obtained are consistent with experimental investigations based on X-ray diffraction and neutron scattering on different tetrahedral glasses and amorphous materials such as silicon and glassy selenium \cite{wilding2006, dyre2006}. In these works, double-peak structures are observed, the two main peaks being found at distances consistent with our model. Daisenberger et al.\ also found a similar $S(q)$ profile for silicon with X-ray diffraction experiments \cite{daisenberger2007, loerting2009}. In their work, they control the system pressure in analogy with our control of the packing fraction. In addition, Mollica et al.~\cite{Mollica2022} used MD simulations of significant complexity to study amorphous ice in silico using a TIP4P/ice force field where the simulated sample reduces to significantly narrow widths. In spite of the vast differences between their model and ours, the results are consistent: they show a $S(q)$ for the oxygen-‐oxygen correlations with a double-peak structure and a behavior at low $q$ that resembles the one we observe at intermediate packing fractions $\phi=0.3$.

\begin{figure}[tb]
\centering
\includegraphics[width=1.0\linewidth]{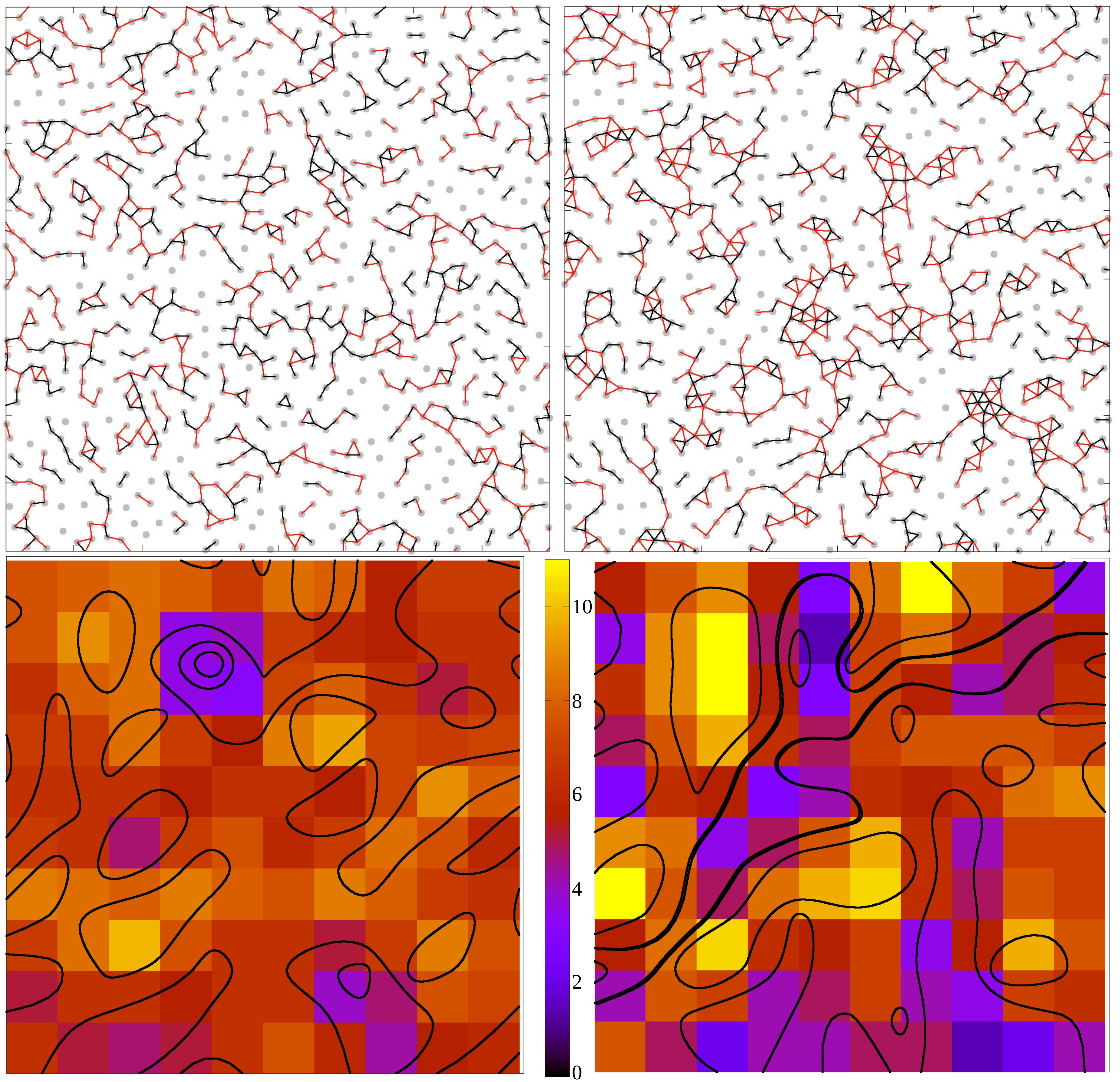}
\caption{\textbf{Snapshots} (top) for $\phi=0.3$ and $n=3$ (left) and $6$ (right). Black (red) lines between particles represent bonds in the first (second) potential well. \textbf{Contour maps} (bottom) of the snapshots colored according to the number of particles per square; contour lines join points at constant density; lines are smoothed for clarity using B-splines. Two particularly long contour lines  of low particle density have been highlighted for $n=6$.}
\label{snapshots}
\end{figure}

To have a  clearer intuition of the structure manifested through $S(q)$ and $g(r)$ at intermediate $\phi$, in Fig.~\ref{snapshots} (above) we show snapshots of a large control surface of two stable configurations with the same density of particles for $\phi = 0.3$ with $n = 3$ (left) and $n = 6$ (right), where bonds in the first (second) well are depicted as black (red) lines. Although in both cases the system is not fully connected into a single cluster, the clusters we see for $n = 6$ spread to larger areas and involve more particles. This is expected since at high $n$, particles can bind to more neighbors and therefore propagate their connections further. Insets in Fig.~\ref{sq} show for all the $\phi$ values investigated the mean number of bonds per particle, as well as the potential energy per particle, $V$/particle, as a function of $n$. There we see how, for $\phi = 0.3$, particles with $n = 3$ present around 2.5 bonds per particle on average whereas this number reaches a value around 4.5 with $n = 6$. The long-range structure we see for $S(q)$ at low $q$ ($\approx 2$) and intermediate $\phi$ can also be observed through the particle configurations. This is what we show in Fig.~\ref{snapshots}(bottom), where the previous snapshots have been coarse-grained by dividing the control surface into squares of side 3.58 ($\approx$ 2$\pi$/$q$, where $q\approx 2$). We assign a color to each square according to the number of particles it contains. Whereas for $n = 3$ (left) we see that colors are mostly uniform, pointing to a uniform local density, for $n = 6$ (right) we clearly see strong local deviations from the average density with high density (light yellow) and low density (violet) squares. This observation points to the emergence of local spatial heterogeneity at high $n$ for the length scale where the plateau in $S(q)$ is observed. Both observations, the emergence of large clusters and heterogeneous local density, are also manifested through the contour lines following constant densities that we show in Fig.~\ref{snapshots} (below). Since our control surface has a $10\times10$ square resolution, we applied a B-spline smoothing method to the contour lines to improve their visibility \cite{MacCallum1986}. Whereas for $n = 3$ (left) contour lines with a given density typically extend to small areas, for $n = 6$ (right) we see some contour lines that prolong to large distances. In particular, we have highlighted two contour lines, corresponding to low density, that cross the entire control surface. Similar snapshots to Fig.~\ref{snapshots} have been observed in the deposition of a monolayer of water on hydrophilic surfaces \cite{brovchenko2008}.

\begin{figure}[tb]
\centering
\includegraphics[width=1.0\linewidth]{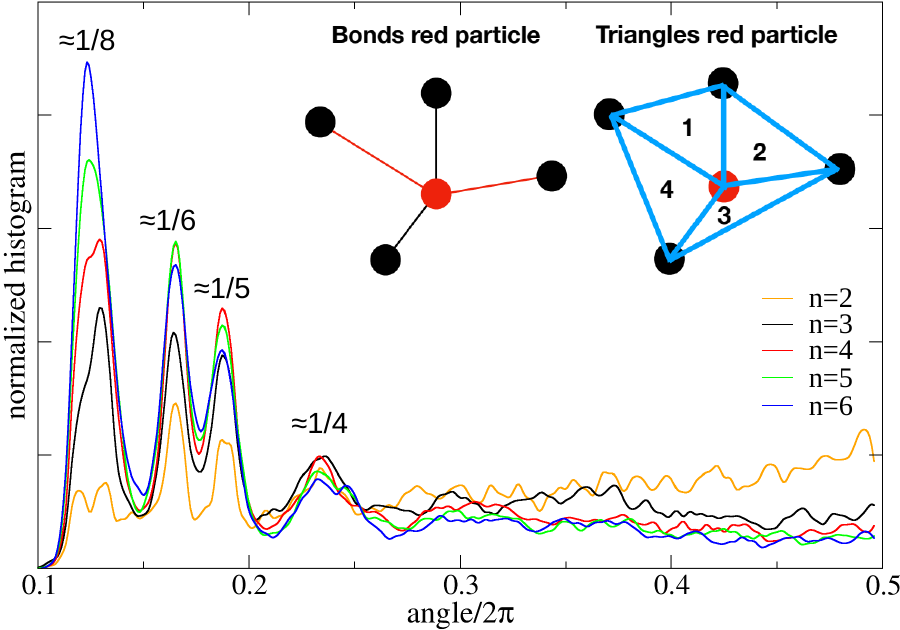}
\caption{\textbf{Angle distribution} shown by a normalized histogram for $\phi$=0.3 as a function of $n$ of the angles associated with the non-overlapping triangles each particle forms with any pair of its adjacent neighbors. Sketches show (left) a generic particle (red) with four neighbors (lines represent bonds as in Fig.~\ref{snapshots}), and (right) the corresponding four blue triangles (numbered from 1 to 4) that the red particle forms with any pair of its adjacent neighbors, from which we obtain the angles to construct the histograms.}
\label{angles}
\end{figure}

{Particle connectivity in our system can be studied under different  approaches, for instance through the use of planar graphs \cite{Bondy-Murty}. We, however, choose to analyze this problem by the local particle orientations, in particular at intermediate $\phi$. We keep} in mind the fact that interaction in our model is isotropic and, therefore, does not contain explicit directionality. Fig.~\ref{angles} presents the local angular distribution according to the construction we show by the sketch. For any given particle  having at least two neighbors (see red particle in the sketch), we construct all the triangles formed by the particle and any pair of its adjacent neighbors (see blue triangles in the sketch). We then collect all the angles of the triangles associated with all the particles in the system having at least two neighbors and study the different normalized histograms corresponding to each $n$ value. We see that all the histograms present four well developed peaks placed, from left to right, around $\pi/4$, $\pi/3$, $2\pi/5$, and $\pi/2$; as well as a rather uniform spectrum at large angles in the form of a long tail. These distributions show the interplay present in our system between local polycrystallinity, expressed by well defined angles, and amorphous structure, given by a long tail at large angles.

The well-defined angles we observe can be typically associated with specific triangular structures (see Fig.~\ref{snapshots} (top)): i) $\pi/4$ and $\pi/2$ could be naturally explained by the presence of square-like structures (with the red particle at a vertex) with a right angle ($\pi/2$) and two angles  ($\pi/4$) defining the diagonal; ii) $\pi/3$ would be associated with equilateral triangles that appear dispersed or as part of a hexagonal structure (with the red particle in the center); and iii) $2\pi/5$ being associated with pentagonal structures. However this idealized interpretation does not take into account the two distances, potential wells, defining the system interaction. These two distances redound to the width of the peaks we observe as well as in their precise location. Although no directional interaction is imposed in our model, the revealed peak structure becomes more apparent when increasing $n$, whereas the corresponding tail at large angles decreases, leading to better defined local polycrystalline structures. This is in clear contrast to the broad and centered  distribution observed for non-overlapping disks randomly deposited when analyzing their Voronoi polygons \cite{hinrichsen1990}.

\section{Conclusions}
We have introduced a minimal athermal model in two dimensions with isotropic interactions and three basic ingredients: a double-well bonding mechanism, an excluded volume, and a maximum coordination number. The model shows that even without any thermodynamical considerations, it is possible to   
produce all the relevant features observed in several amorphous substances such as long-range order, spatial heterogeneity, and local polycrystallinity. 
{We have explored for this model packing fractions ranging from dilute systems to values close to the random packing fraction of equal non-overlapping disks. We have also investigated the effect of valence on our particles, varying their coordination number from 2 to 6.
}

Our results are applicable to a  variety of systems including 2D liquid water and ice \cite{meyer1999,koga2000,johnston2010,chen2016,ma2022,negi2022,Mollica2022,brovchenko2008,bartels2012}, silicon \cite{sastry_silicon,treacy2012,daisenberger2007,loerting2009}, and different glassy and amorphous solids \cite{dyre2006,loerting2009,wilding2006}. {In summary, we have shown that for intermediate packing fractions and high coordination numbers (5 and 6), the radial distribution function of our model shows a similar short- and long-range structure to that reported in more complex 2D models of amorphous water and ice \cite{koga2000,negi2022} as well as in experiments on amorphous silicon \cite{treacy2012}. Again, at intermediate packing fractions, the structure factor of our system resembles that observed in simulations of amorphous ice in confined geometries \cite{daisenberger2007}, whereas at high packing fractions it is similar to those reported in experiments on silicon \cite{wilding2006} and glassy selenium\cite{dyre2006}. In addition, at intermediate values of the packing fraction and high coordination number, we have noted spatial heterogeneities similar to those observed in the deposition of a monolayer of water on hydrophilic surfaces \cite{brovchenko2008}.} These results support the hypothesis that a double-well potential, {together with a maximum coordination number}, may be sufficient to give rise to the coexistence of different types of local competing structures emerging in  materials displaying polyamorphism \cite{buldyrev2002,wilding2006,Rosu-Finsen2023,Mollica2022,shi2020}.

{Two possible extensions of this model are: (i) its implementation in 3D and (ii) the ability to solve the corresponding inverse problem, i.e. how a priori to tune the model parameters to reproduce the properties of a previously chosen amorphous system. 
However, we consider that, due to the phenomenological nature of our model, the inverse problem would need a systematic investigation.
}
{Finally, a straightforward expansion of our current model could include the effect of external perturbations leading to surface deformations. For instance, external pressures, whether positive or negative, might be simulated by implementing a variable system surface. Such an approach is expected to produce structural patterns such as those observed in high-pressure amorphous ices. The model can also simulate fractures, if allowing bonds to be broken when stretched beyond a set distance.
}

\begin{acknowledgments}
B.E. acknowledges support from grants PCIN-2017-098, PTA2020-018247-I and PID2020-118974GB-C21 from the Spanish Ministerio de Ciencia e Innovaci\'on. 
{J.H.E.C. and C. I. S. D. acknowledge support from the Spanish Ministerio de Ciencia, Innovaci\'on y Universidades through grant PID2024-160443NB-I00.}
S.R.-V. acknowledges support from the European Commission through the Marie Skłodowska-Curie Individual Fellowship 840195-ARIADNE.
This project has received funding from the European Research Council (ERC) under the European Union’s Horizon Europe research and innovation programme ERC-AdG-2022 (GA No. 101096293). Views and opinions expressed are however those of the author(s) only and do not necessarily reflect those of the European Union or the European Research Council Executive Agency. Neither the European Union nor the granting authority can be held responsible for them.
\end{acknowledgments}

\section*{Data Availability Statement}
The data that support the findings of this study are available from the corresponding author upon reasonable request.

\bibliography{Biblio}

\end{document}